# Particle migration of suspensions in a pressure-driven flow over and through a porous structure

Parisa Mirbod[a*], Nina C. Shapley[c]

[a] Department of Mechanical and Industrial Engineering, University of Illinois, Chicago, IL
[c] Department of Chemical and Biochemical Engineering, Rutgers, The State University of New Jersey, Piscataway, NJ
[*] Corresponding Author: Email address for correspondence: pmirbod@uic.edu (P. Mirbod)

**Abstract**

Laboratory experiments were conducted to study particle migration and flow properties of non-Brownian, non-colloidal suspensions ranging from 10% to 40% particle volume fraction in a pressure-driven flow over and through a porous structure at low Reynolds number. Particle concentration maps, velocity maps and corresponding profiles were acquired using a magnetic resonance imaging (MRI) technique. The model porous medium consists of square arrays of circular rods oriented across the flow in a rectangular microchannel. It was observed that the square arrays of the circular rods modify the velocity profiles and result in heterogeneous concentration fields for various suspensions. As the bulk particle volume fraction of the suspension increases, particles tend to concentrate in the free channel relative to the porous medium while the center-line velocity profile along the lateral direction becomes increasingly blunted. Within the porous structure, concentrated suspensions exhibit smaller periodic axial velocity variations due to the geometry compared to semi-dilute suspensions (bulk volume fraction ranges from 10% to 20%) and show periodic concentration variations, where average particle concentration is slightly greater between the rods than on top of the rods. For concentrated systems, high particle concentration pathways aligned with the flow direction are observed in regions that correspond to gaps between rods within the porous medium.

## 1. Introduction

Flow and transport over and within porous media are ubiquitous in many natural and industrial processes. Researchers have investigated flow over sediment beds [1], coral reefs and submerged vegetation canopies [2], crop canopies and forests [3], endothelial glycocalyx of blood vessels [4-6], flow over carbon nanotubes (CNTs) [7], and polymer brushes [8], to name a few. Suspension flows, on the other hand, are present in many applications including biological systems (red and white blood cells), food processing, costume products, and waste slurries [9]. However, we poorly understand the coupled flow and the behavior of particles over and inside porous structures.

The velocity and concentration profiles for two-dimensional (2D) steady state flow of non-Brownian suspensions in a channel with smooth walls have been examined both theoretically and experimentally in various studies [10-26]. Shear-induced particle migration has also been examined computationally using discrete particle simulation approaches, including Stokesian dynamics (SD), the immersed boundary method (IBM), and the force coupling method (FCM)[27, 28]. It is also known that the complex behavior of a suspension in smooth geometries depends on the properties of the suspending fluid and the microstructure of the dispersed phase.

From an experimental point of view, optical measurement techniques are now routinely used in experimental fluid mechanics to investigate pure Newtonian fluids, dilute and concentrated suspensions. For instance, Laser-Doppler velocimetry (LDV) [29-32] and Particle Image Velocimetry (PIV) [33-45] are increasingly used to visualize flow behavior in smooth geometries, either channel or tube, and in both laminar and turbulent regimes. A related, non-optical scattering method, Ultrasound Imaging Velocimetry (UIV), has been used by [46] to analyze turbulent particle-laden flows at moderately high particle volume fractions. PIV has also been employed extensively to examine pure Newtonian flow over and through porous structures [47-51].

The focus of this work is to address the coupling behavior of suspension flow and a model porous medium. Specifically, we employ magnetic resonance imaging (MRI) measurements to examine suspension systems ranging from 10% to 40% particle volume fraction. MRI is a non-optical and non-scattering measurement method that provides comprehensive, full-field, two-dimensional (2D) maps of the particle distribution and steady state flow field in a three-dimensional (3D) geometry, even in visually opaque and highly concentrated suspensions [13, 15, 16, 52-69]. Therefore, MRI offers advantages beyond those of widely used optical techniques, such as LDV and PIV, namely, the ability to obtain particle concentration data and to access optically opaque systems. The obtained measurements of the various cross-sectional particle concentration and velocity distributions provide key information leading to deeper understanding of the suspension flows and particle motion over and inside a porous structure.

In this study, we provide quantitative evidence and new insights into particle migration and flow of non-Brownian, non-colloidal suspensions ranging from 10% to 40% particle volume fraction over and through a structured surface in a pressure-driven flow. In particular, we focus on revealing the flow behavior and particle concentration distribution, and describing the underlying physical processes leading to these changes in the presence of a structured surface. Herein, the migration of particles due to inertial forces are negligible, meaning that the motion of the suspension is in the Stokes flow regime that occurs in many applications of suspension processing. The channel bottom surface is covered by a porous media model with known physical properties (i.e., porosity of 0.9, permeability of 7.93 x $10^{-7}$ m$^2$), and thickness of 0.5 cm, while there is a free flow region with a thickness of 0.2 cm above the porous structure. We characterize the impact of suspensions and their interaction over a porous media model and compare the data with both the pure suspending fluid flowing over a porous media model and suspensions in a smooth channel. The experimental setup is described in section 2, the results are presented and discussed in section 3, and concluding remarks are presented in section 4.

## 2. Experimental setup and procedure

### 2.1. Experimental apparatus, porous media model, particles, and fluid

The experimental setup in figure 1 (a, b) contains a channel with a length $L = 66$ cm, width $W = 2.5$ cm, and total height $H = 0.7$ cm. The porous media model is 3D printed and consists of circular rods in a square array aligned perpendicular to the flow direction. The rods have a diameter of $d = 0.15$ cm and the space between the sides of the rods is 0.27 cm. The porous medium occupies the bottom 0.5 cm of the channel height and there is a free flow region above the porous medium with a thickness of 0.2 cm. The porosity, $\varepsilon$ of this model is then calculated to be 0.9 and the resulting permeability, $K$ based on the rod diameter and the porosity is found to be 7.93x$10^{-7}$ m$^2$. These analyses are based on theoretical formulas reported in [70] applied to the section containing the cylindrical pillars. Considering the porosity ($\varepsilon$) and spacing distance between the rods (S) according to the values in Table 1, and the diameter of the pillars ($d$), we calculated the porosity using $1 - \varepsilon = \frac{\pi d^2}{4S^2}$. We then defined the permeability of each porous structure considering the solid pillars in a one-dimensional square arrangement by employing $\frac{K}{d^2} = \frac{0.16\left[\frac{\pi}{4\varphi} - 3\sqrt{\frac{\pi}{4\varphi}} + 3 - \sqrt{\frac{4\varphi}{\pi}}\right]}{\sqrt{1-\varphi}}$, where $\varphi$ is the solid volume fraction. While these calculations have been reported in our prior works (e.g., [43, 45, 71]), verifying the exact value of the permeability is the subject of our current investigations. Table I reports the parameters used in this study, including the length, porosity, and height of the porous media model $l, \varepsilon, h_p$, the channel gap and width $H, W$, the diameter of the rods, $d$ and the spacing between the rods, $S$. Note that the coordinate system starts at the top of the rods, i.e., y=0 is located at the porous model interface.

More details have been reported in [43]. The flow loop uses a peristaltic pump (MasterFlex L/S, Cole-Parmer) with an additional pump head (Easy Load II, Cole-Parmer) and flexible silicone L/S 15 pump tubing, yielding nominal flow rates ranging from 20 to 1000 ml/min. The pump tubing feeds into a pulse dampener (Masterflex Pulse dampener, Cole-Parmer) which helps to reduce any pulsating from the pump. The resulting flow is smooth and at a constant speed where it enters the flow channel. The actual flow rate throughout this study has been measured by collecting the suspension in a prescribed time interval at the outlet. The pump draws the suspension from an inlet reservoir and brings it to the pulse dampener. After exiting the pulse dampener, the suspension flows through flexible silicone L/S 15 pump tubing of inner diameter 3/16 inch (0.476 cm) and length of 38 cm. It then expands into a smooth rectangular channel with a height of 0.7 cm, width of 2.5 cm, and length of 21.5 cm before flowing through the porous medium. The porous media model has a length of 28.5 cm, and then there is another smooth rectangular channel section of 16 cm length, where the flow then contracts to an outlet tube of 33 cm length and 1/2 inch inner diameter (1.27 cm). After flowing through the flow channel and outlet tube, the suspension is collected in an outlet reservoir, which was periodically poured into the inlet reservoir to recirculate the suspension. For the measurements presented here, the volumetric flow rate $Q$ was set at a constant nominal value of 70 ml/min, and the masses of catch samples collected from the outlet tube confirmed that the actual volumetric flow rate of 70 ml/min was achieved, for bulk particle volume fractions up to and including 40%.

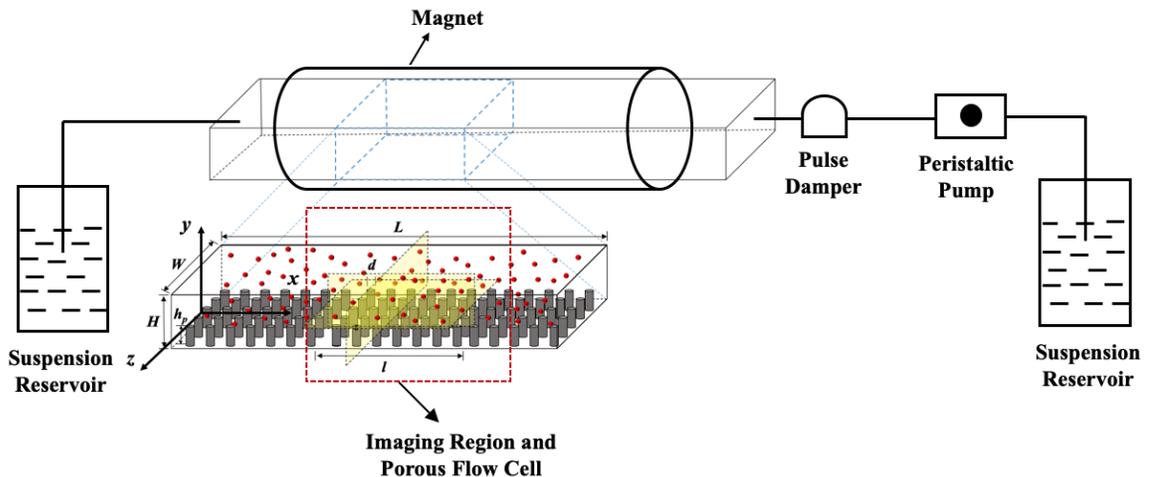

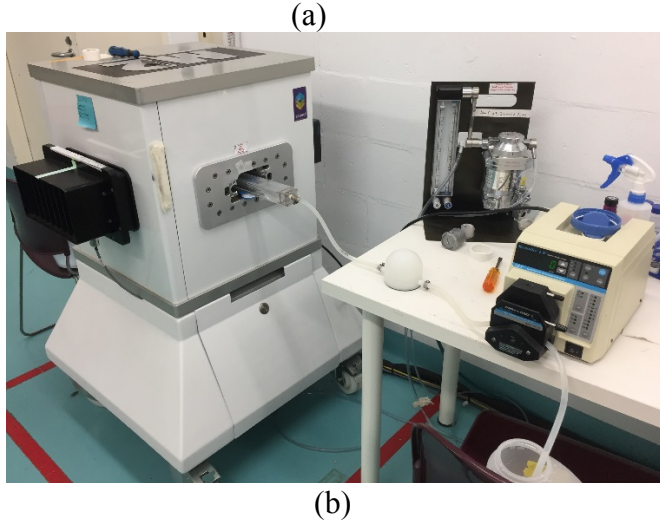

**FIGURE 1.** (a) Sketch of the experimental setup including the geometry of the channel with the porous texture for exploring suspension flows over a porous structure model using MRI measurements and (b) a photo of the experiments performed at Rutgers Molecular Imaging Center. The values are reported in Table I. The coordinate system starts at the top of the rods, i.e., y = 0 located at the porous model interface.

TABLE I. Experimental parameters for various suspensions flowing through and above a porous media model.

| $\varepsilon$ | $h_p$ (cm) | $l$ (cm) | $d$ (cm) | $H$ (cm) | $L$ (cm) | $W$ (cm) | $S$ (cm) |
|---|---|---|---|---|---|---|---|
| 0.9 | 0.5 | 28.5 | 0.15 | 0.7 | 66 | 2.5 | 0.27 |

The suspensions contain spherical polymethyl methacrylate (PMMA) particles (Lucite 41, Lucite International, Cordova, TN). The particle distribution is broad and monomodal, with a mean diameter of 96 μm and standard deviation of 25 μm. The particle density is approximately 1180 kg/m$^3$, with some variation due to trapped monomer bubbles during manufacturing by emulsion polymerization. In order to achieve density matching and prevent buoyancy effects a solution of glycerin (73 wt%) and water (27 wt%) with a density of 1180 kg/m$^3$ has been used. The Newtonian solvent viscosity, $\eta_0$, was measured with a parallel plate rheometer Kinexus Rheometer (KNX2100, Malvern Instruments) and found to be 0.025 Pa.s at 25°C. For each experiment, the suspending fluid was well mixed and then particles were added. The resulting suspension was put into a vacuum chamber and subjected to a pressure of -30 inHg, and left for an additional hour to remove entrained air bubbles from the suspension. Only very minimal loss of water was detected in the vacuum pump trap, resulting in an insignificant change in the suspending fluid viscosity due to degassing. For the MRI measurements of each suspension, the flow cell was loaded initially with the pure suspending fluid in order to acquire calibration images, and then emptied. Next, the flow cell was filled with the suspension at a low flow rate of 50 ml/min, to avoid trapped air bubbles. The pump flow direction was oscillated for at least 5-10 minutes at flow rates ranging from 50-200 ml/min in order to make the starting particle concentration distribution as uniform as possible and to minimize loading effects following the work reported by [72]. Once the pump was started, the flow was left to stabilize through the flow loop for 20 minutes to reach a nearly steady state condition before any data was collected. The flow was not observed to vary over the duration of each velocity measurement, which was approximately 1 minute.

*2.3. Experimental Procedure and data analysis*

To explore the impact of particle concentration on the flow patterns over and through the porous media model, various suspensions with bulk particle volume fraction $\phi_b$ =10%, 20%, 30%, and 40% were tested. Each experiment was performed where the flow was at an apparent steady state over the duration of the measurements and also where concentration variations from rod to rod along the flow direction were not observed over the 5 cm × 5 cm field of view of measurement, which included at least 12 rods along the flow direction. The suspending fluid closely matched the particle density of 1180 kg/m$^3$ so that particle settling was not detected over the experiment duration. Over a long period of rest, such as overnight, rising of the particles was observed, since the particles had a slightly lower density than that of the suspending fluid. However, during the experimental runs, particle settling was also counteracted by frequent mixing of the inlet reservoir and the shearing motion present in the flow itself, which is known to promote resuspension [73].

In this work, we were particularly interested in flows with low Reynolds number, allowing the suspension to be in the Stokes flow regime. To characterize the flow, we considered three different Reynolds numbers due to the complexity of the flow; 1) the Reynolds number based on the PMMA particle size, $Re_p = \frac{4\rho_f a^3}{3\eta_0 L^2}|U_{max}|$, where $\rho_f$ is the density of the fluid, $\eta_0$ is the viscosity of the Newtonian fluid, $L$ is half width of the free-flow region, $a$ is the average radius of the PMMA particles, and $U_{max}$ is the streamwise maximum velocity within the flow channel [20, 21], 2) the Reynolds number based on the porous media, defined as $Re_L = \frac{\rho_f d U_b}{\eta(\phi)}$, where $d$ is the diameter of the rods making up the porous media, and $U_b$ is the streamwise bulk velocity $\left(U_b = \frac{1}{H_T}\int_{-H}^{2L} u(y)dy\right)$, and $\eta(\phi)$ is the suspension viscosity proposed by Krieger [74], and 3) the suspension Reynolds number which is based on the half of the free-flow region as the characteristic length. The related equation for the suspension Reynolds number can be given by $Re_S = \frac{\rho_f L U_b}{\eta(\phi)}$. We then found $Re_p \sim O(10^{-6})$; therefore, the inertia of the particles can be neglected. The porous and suspension Reynolds number also computed as $Re_L \sim O(10^{-1})$ and $Re_S \sim O(10^{-1})$, respectively. It should be noted that the Stokes number was calculated to be $St \sim O(10^{-5})$ and the Brownian Péclet number, is $O(10^9)$ for the range of shear rate used in this work [59]. The Stokes number is defined as $St = \frac{m_p \dot{\gamma}}{3\pi \eta_0 d_p}$ and the Péclet number as $Pe = \frac{3\pi \eta_0 d_p^3 \dot{\gamma}}{4kT}$ where $m_p$ is the mass of the particles, $d_p$ is the diameter of the particles, $kT$ is the thermal energy of the solvent, and the shear rate $\dot{\gamma} = \frac{U_{max}}{L}$. Therefore, the suspensions studied here are properly characterized as non-Brownian, non-colloidal, and flowing at low Reynolds number. We also indicate that to ensure there are enough particles across the gap, the ratio of the particle diameter and the channel gap is very small ($\approx 0.012$); therefore, the suspension can be considered as a continuum medium. We also compared the ratio of the gap and the length of the channel to the guideline proposed by [17, 20, 21] for the particles to reach a fully developed region. For 40% bulk particle concentration, the estimated fully developed length in the smooth region is around 10 m. In all the measurements, however, we selected the length of the channel to ensure that the secondary flows do not have a dominant impact on the measurements as reported in [75, 76]. For our experiments, the length at which the secondary flow has a significant impact was estimated to be 70 cm. We therefore designed the experimental setup to minimize the impact of secondary flows (in particular for higher concentrations) while the suspension flow might be still developing. It is noteworthy that the porous media has a three-dimensional (3D) geometry resulting in a 3D suspension flow inside the channel studied here. We then examined the nearly fully developed flow for our proposed porous model in detail in Section 3.

*2.4. Flow visualization technique*

The velocity and concentration measurements were conducted using MRI measurements on both the pure suspending fluid and suspensions with bulk volume fractions ranging from 10% to 40%. The 3D concentration and velocity measurements were accomplished on an Aspect Imaging M2 Compact High-Performance MRI (1 Tesla) microimaging system with maximum imaging gradients of 400 mT/m, for proton (H1) MRI at the Rutgers Molecular Imaging Center. The MRI is computer controlled by a SGI workstation running Aspect Imaging "NRG" software. We used MATLAB and other imaging processing programs such as ImageJ for analysis of acquired data. For this MRI instrument, in-plane resolution of 200-400 μm pixel width and out-of-plane slice thickness of 0.5-2 mm are typical values. For the images presented here, 390 μm pixel width and 1 mm slice thickness were used, in order to obtain sufficient signal to noise ratio required for distinguishing spatial variations in particle volume fraction and axial velocity. Note that our MRI measurements were centered around an axial distance of 23.2 cm from the beginning of the porous region, with a field of view of 5 cm × 5 cm. Since the total length of the porous region is 28.5 cm, the imaging volume is near the end of the porous region but does not include the end.

For the axial velocity measurement, we employed the phase encoding method, where the scan sequence was spin-echo diffusion weighted imaging (SEDWI). The phase of the MRI signal was obtained from each pixel of the image. The phase acquired when the flow was stopped was subtracted from the phase measured during flow, and the difference is proportional to the local velocity [58, 77, 78]. The phase encoding measurements are obtained on various y-z planes near the center of the imaging region, evenly spaced on top and in between the rods. We processed all the raw data from phase encoding measurements in different planes using in-house MATLAB programs to give velocity maps and provide the velocity distribution in various channel cross-sections using the method reported in [79]. The velocity map is stored in a matrix and profiles can be calculated based on average velocities across the different sections. The suspension flow rate can be further computed by multiplying the flow average velocity by the channel area.

We used the spin-echo pulse sequence (SE 2D) for concentration measurement of the data obtained from MRI. This pulse sequence incorporates a slice selective 90° pulse followed by one or more 180° refocusing pulses. The image contrast by the spin-spin ($T_2$) relaxation time difference between the solid particles and the suspending fluid represents the binary suspension system. The high intensity signals in the images, shown bright on a gray scale (see Fig. 2), correspond to low particle concentration. This is because the particles do not contribute to signal intensity due to their short T$_2$ relaxation time. Particle concentration distribution images are also taken on various slices in the same locations as for the velocity distribution images. The intensity measurements are obtained on various y-z and x-y planes evenly spaced on top and in between the rods. We also take the intensity images on three vertical-facing slices (x–z planes) at y = $h_p$- 0.5 mm (top 1 mm of rod), y = $h_p$ -1.5 mm (middle of rod), and y = $h_p$+ 0.5 mm (above rod), where the slice thickness is 1 mm and the gap between slice centers is 1 mm.

The concentration quantification process (i.e., the conversion from image signal intensity to particle volume fraction) resembles that employed by [59]. As a first step, we have to remove the image artifacts produced due to the radio frequency (rf) field nonuniformity. To do this, images are acquired from the pure suspending fluid to serve as calibration images. The pure fluid generates spatially uniform signal strength where the resulting image nonuniformity largely captures the magnetic field artifacts that can also be observed in the suspension images. It was found that the calibration procedure reduced the typical image intensity variation relative to the mean value by up to 20%. The calibration is performed with MATLAB and Excel programs. Once the artifacts are removed, the signal intensity of each pixel is directly proportional to the fluid volume fraction at that location. We further characterized particle flux in the channel over and through the porous media model. Further details of the procedure can be found in [67]. Fig. 2 represents a sample of the intensity images taken from the MRI technique, before any concentration analysis was performed, showing the intensity distribution over and through the porous media model for $\phi_b$ =30% suspension at different coronal planes including inside the free-flow region, the porous region, and at the suspension-porous interface. The dark regions clearly indicate that particle migration has occurred. Fig. 2(a) shows migration of particles in the free-flow region that can be clearly observed, where a large width dark band generated in the middle of the free-flow region runs parallel to other dark bands close to the channel wall. The dark regions produced due to particle migration between the rods can also be faintly observed in Fig. 2(b) and (c). Figure 3 represents an example of axial intensity images obtained for the pure suspending fluid and $\phi_b$ =40% suspension at different axial planes on top of and in between the rods. While the suspending fluid images (a,b) appear to be quite uniform, intensity variations are clearly apparent in the suspension images after flow (e,f), where the top of the image appears to be darker, corresponding to higher particle concentration.

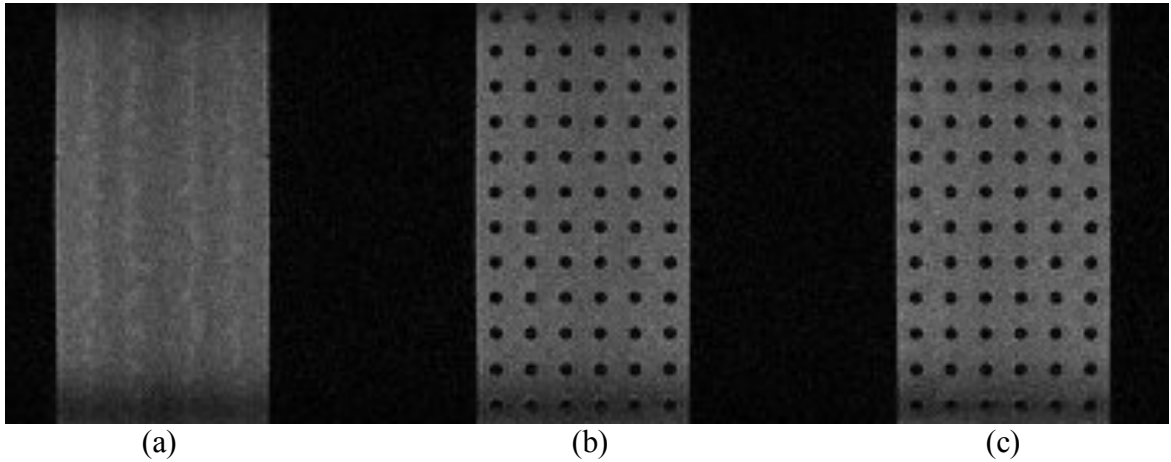

(a)                                  (b)                                 (c)

**FIGURE 2.** Intensity images showing particle distribution over and through the porous media model for $\phi_b$ =30% suspension. (a) flow in the free-flow region (y=$h_p$+ 0.5 mm = 5.5 mm), (b) flow at the suspension-porous interface (y=$h_p$- 0.5 mm = 4.5 mm), and (c) flow in the middle of the rods (y=$h_p$-1.5 mm = 3.5 mm). The signal intensity is proportional to the fluid volume fraction, so that low intensity (dark) pixels correspond to high local particle concentration. The length of the flow cross-section (bright region) is 5 cm and the width is 2.5 cm. MRI scan parameters: spin-echo 2D pulse sequence, TR=2 s, TE=13.7 ms, field of view=5 cm × 5 cm, slice thickness=1 mm, 128 × 128 pixels.

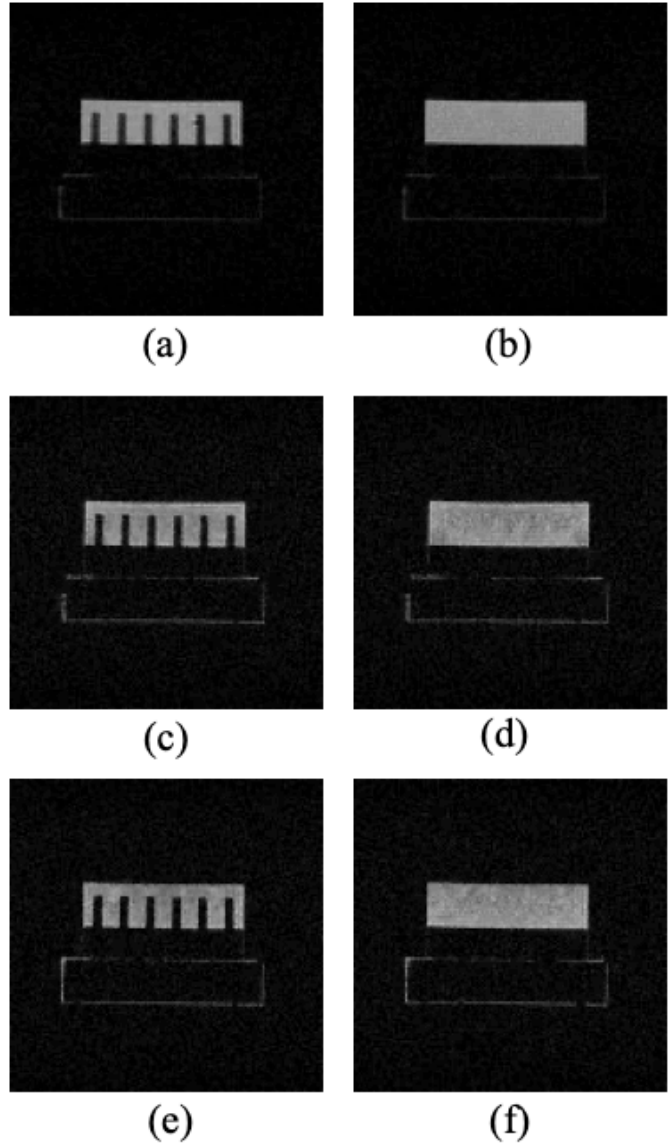

**FIGURE 3.** Intensity images showing particle distribution over and through the porous media model for (a,b) pure Newtonian suspending fluid, (c,d) suspension flow ($\phi_b = 40\%$) before, and (e,f) after 30 min of the flow for Q=70 ml/min. (a,c,e) flow on top of the rods, and (b,d,f) flow in between the rods. The height of the flow cross-section (bright region) is 7 mm and the width is 2.5 cm. MRI scan parameters: spin-echo 2D pulse sequence, TR=2 s, TE=13.7 ms, field of view=5 cm × 5 cm, slice thickness=1 mm, 128 × 128 pixels.

## 3. Results and discussion

Fig. 4 shows velocity maps of the axial velocity component derived from phase encoding scans for the pure suspending fluid and various flow concentrations ranging from 10% to 40% for flow on top and between the rods located at x=23.2 ± 0.11 cm (1.1 mm image separation, where images of 1 mm thickness alternate between being centered on the rods and halfway between the rods) in the axial direction (y-z plane). The graphs show that the locations of velocity peaks vary with bulk particle concentration, both in the free-flow and the porous region. For pure fluid and bulk concentrations of 10% and 20%, velocity maxima are observed in the centers of the gaps between rods, without noticeable variation along the spanwise direction from rod to rod. In contrast, for the higher bulk concentrations of 30% and 40%, significant variation is observed along the spanwise direction, where the velocity distribution between the porous media model shows peak velocity values near the channel side walls that decays towards the channel centerline. The velocity contours for 40% concentration suspension are shown in Fig 4(i) for flow on top and in Fig. 4(j) for flow between the rods. These contours resemble those of 30% concentration tests; however, the flow velocity near the centerline both in the free flow and the porous region decays with increasing particle concentration.

Such an effect has also been reported previously in [67] for suspensions through an asymmetric bifurcation of a smooth, low aspect ratio rectangular channel, but differ from other works. For example, Lyon and Leal (1998) considered high aspect ratio channels where the wide dimension can be completely ignored. In this study, the low cross-sectional aspect ratio (3.57:1) of the channel results in non-negligible behavior of the wide dimension in the flow behavior. In the prior work of [67], these two-peak velocity profiles were associated with particle concentration inhomogeneities that developed in the low aspect ratio rectangular channel, of very similar aspect ratio to that of the system studied here. In our current flow cell, there is an opportunity for particle migration to occur in the smooth channel, of 21.5 cm length, upstream of the porous medium. We also observe that regions of high-volume fraction and low shear rate are consistent (see e.g., Fig 6(a, c) and Fig. 7(g, h)). In the velocity images shown in Fig. 4 (for pure suspending fluid and $\phi_b$=0.4 at the same location), it can be clearly observed that the difference between the Newtonian and suspension velocity profiles increases with bulk particle concentration, and therefore likely arises from the tendency of the particles to redistribute into a nonuniform configuration in flow. This effect was observed for the 40% concentration suspensions even after 30 min. The observed behavior is consistent with the concept of the particle stress balance for the pressure-driven flow of suspensions proposed in the suspension balance model by [17, 22]. Specifically, the need for the suspension normal stress to be constant across the directions perpendicular to the mean motion, where the suspension is nearly at steady state and nearly at a fully developed state, results in migration of particles throughout the channel and (in this study) inside the porous region in inhomogeneous ways. It should be noted that the particle size and shape and the thickness, properties, and the 3D structure of the porous media model can play a significant role. These are the subject of our current investigations.

Another considerable difference between the Newtonian and suspension velocity field is the velocity variation with height. For the pure suspending fluid, the maximum velocity values occur in the free-flow region. This is particularly visible between the rods (Figure 4, right hand side images). As the suspension concentration increases, the velocity appears to be more of a plug flow with height. One additional observation is that the difference between the velocity maps on top of the rods and between the rods decreases with concentration. For the pure suspending fluid, maximum velocity values clearly occur on top of the rods with clearly lower velocity values at most locations between the rods. There is some degree of contraction-expansion flow as the cross-sectional area varies periodically in the porous region. In contrast, for the 40% bulk concentration suspension, the velocity maps on top of the rods and between the rods have similar velocity values at each location, including similar regions where the velocity is zero or very low because of the presence of the rods. It appears that concentrated suspensions show a much lower amplitude of contraction-expansion and periodic variations in the porous region than do dilute suspensions and the pure suspending fluid. Although there are some mismatched pixels where the phase of the signal was not correctly captured due to the presence of velocity fluctuations, these general trends are clearly observed.

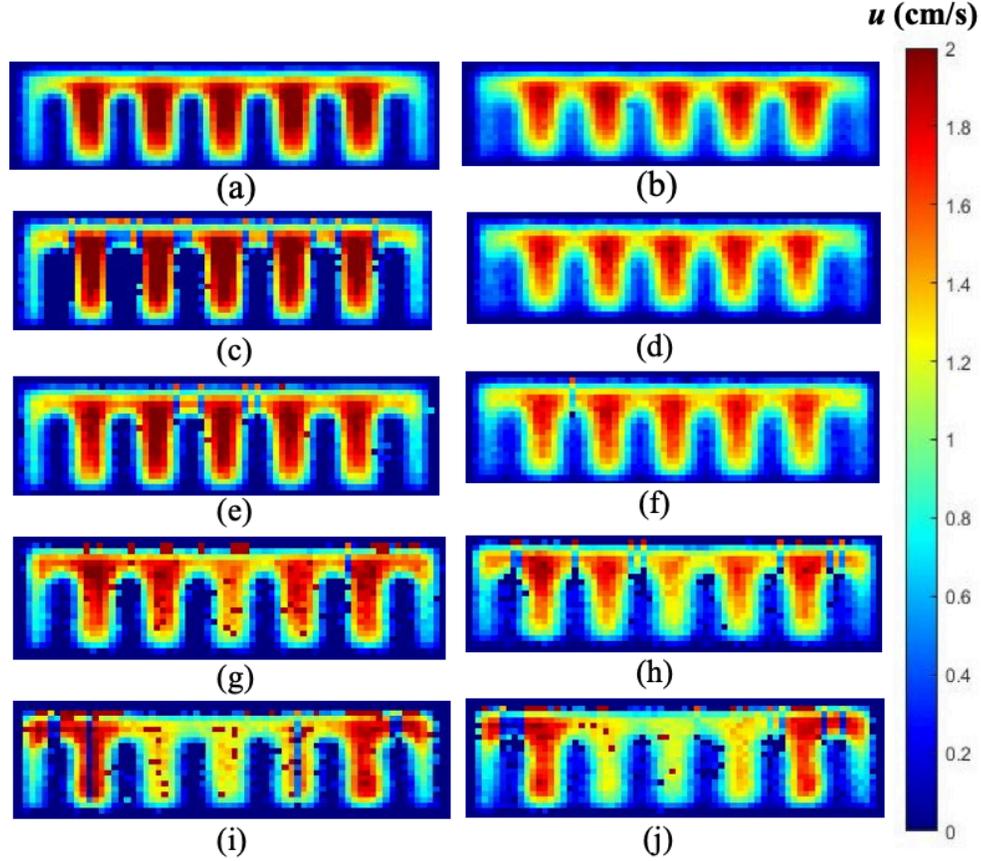

**FIGURE 4.** Axial (streamwise) velocity maps ($u$) (cm/s) viewed in the $y$-$z$ plane (Q=70 ml/min), for (a,b) pure Newtonian suspending fluid, (c,d) $\phi_b$=10%, (e,f) $\phi_b$=20%, (g,h) $\phi_b$=30%, (i,j) $\phi_b$=40%. (a,c,e,g,i) flow on top of the rods, and (b,d,f,h,j) flow between the rods, at distance $x$=23.2 $\pm$ 0.11 cm after the beginning of the porous medium. The height of the flow cross-section (region shown) is 7 mm and the width is 2.5 cm. MRI scan parameters: spin-echo diffusion weighted imaging pulse sequence, TR=500 ms, TE=23 ms, field of view=5 cm x 5 cm, slice thickness=1 mm, diffusion pulse separation $\Delta$=10 ms, diffusion gradient duration $\delta$=2 ms, diffusion gradient =61.2 mT/m, B value=10, 128 × 128 pixels, diffusion gradients in the slice gradient direction, 1 excitation, scan time=64 sec.

Fig. 5 (a, b) show the streamwise ($x$ direction) velocity profiles on the center-line of the channel along the channel height ($y$ direction) on top of the rods and between the rods, respectively. Fig. 5(c) presents the streamwise ($x$ direction) velocity profiles along the channel height ($y$ direction), obtained by averaging both the on top and between the rods' planes for the pure suspending fluid and various suspensions taken on the center-line of the channel. While the Newtonian fluid flow shows a nearly parabolic profile in the free flow region followed by the flow through the rods consistent with those reported by [6, 43, 51] across the *y* direction, by adding particles the velocity values decay in the middle of the channel along the *y* direction. The maximum magnitude of velocity, $U_{max}$, occurs close to the interface for lower concentrations. This is because of the resistance in the flow caused by the existence of the rods for the Newtonian fluid and for very dilute suspensions. As reported in our prior work for both pure Newtonian fluid and very dilute suspensions (i.e., 1%, 3%, and 5%) [59], this effect depends on the thickness and the properties of the porous structure. Also, as discussed above, it can be seen that for dilute suspensions ($\phi_b \leq 0.1$) passing over and through the rods, the flow inside the free-flow region has similar flow physics to that of the Newtonian suspending fluid whereas the velocity decreases inside the rods. However, by increasing the suspension concentration up to $\phi_b = 0.4$, the streamwise velocity inside both the free-flow region and the rods decreases and becomes increasingly blunted. The decrease in center-line velocity and change of profile shape to more of a plug flow could occur because the particles rearrange to try to equalize the pressure between the free-flow and porous regions and minimize the total resistance where they flow through the channel. It is worth noting that since adding the rods results in three-dimensional (3D) flow inside the channel, it is important to examine the spanwise velocity profiles and average them with Fig. 5 to provide information on the impact of rods in the flow. In addition, the thickness of the rods, the 3D structure of the rods and the suspension properties all impact the flow analysis. These are the subject of our current and future investigations.

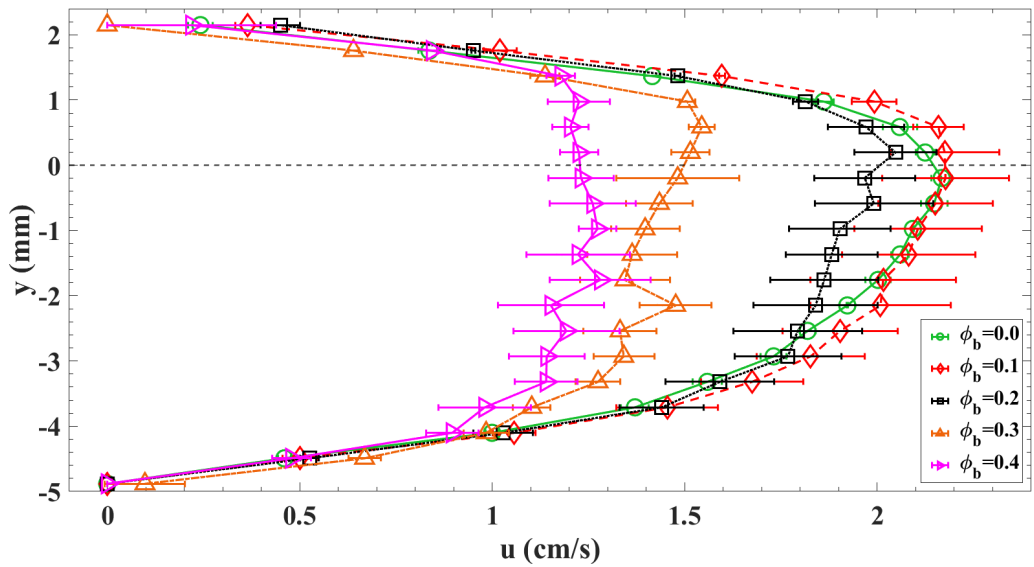

(a)

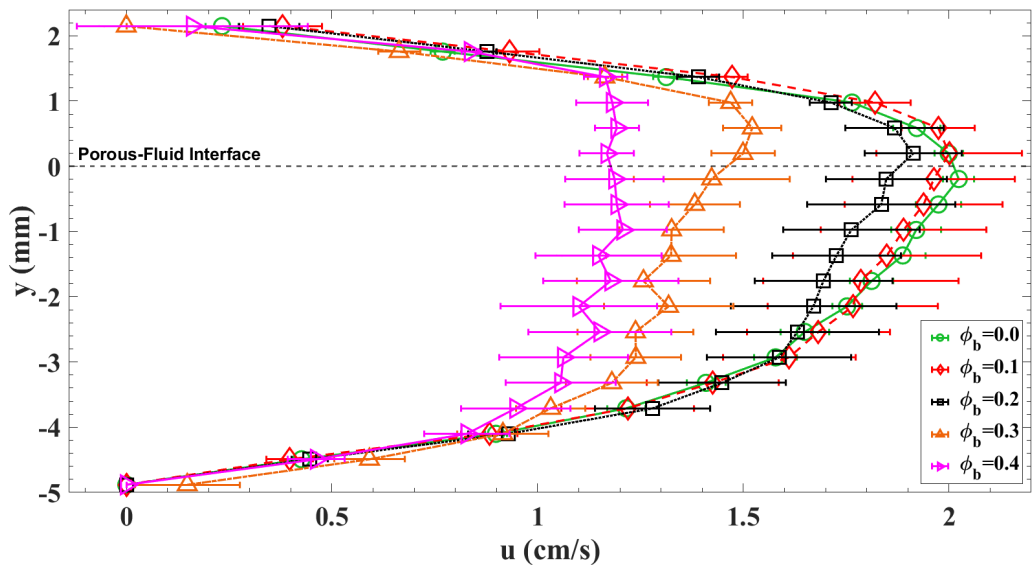

(b)

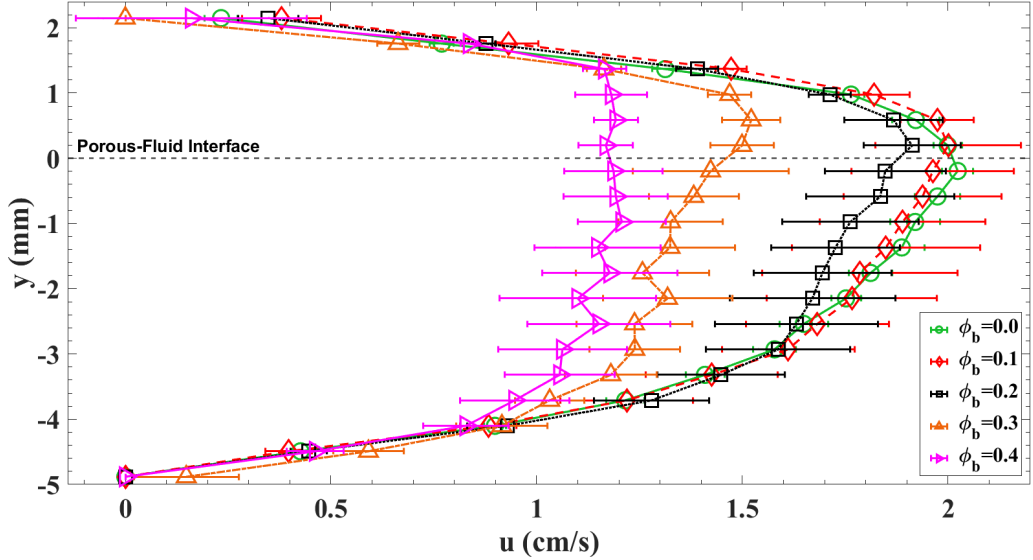

(c)

**FIGURE 5.** Streamwise ($u$) velocity profiles along the channel height ($y$ direction) obtained from the MRI axial velocity maps shown in Figure 4 for (a) flow on top of the rods, (b) flow between the rods, and (c) averaged values of both on top and between the rods, averaging over the 3-4 adjacent pixels on the center-line of the channel for the pure suspending fluid and various bulk concentrations ranging from 10% to 40%. The data on the rods and between the rods of the porous structure were averaged. The volumetric flow rate was $Q = 70$ ml/min for all cases.

The axial, coronal, and sagittal concentration slices (i.e., y-z, x-z, x-y planes), based on intensity images such as those shown in Figures 2 and 3, are also taken at different heights and various locations in the channel. We chose slices at (y = $h_p$ - 0.5 mm (top 1 mm of rod), y = $h_p$ - 1.5 mm (middle of rod), and y = $h_p$ + 0.5 mm (above rod), at x=23.2 $\pm$ 0.11 cm in the y-z plane, and at z=0 (center line) $\pm$ 0.11 cm in the x-y plane to represent the concentration distribution. Figure 6 shows particle concentration distribution maps for $\phi_b = 40\%$ in all three directions for flow on top and between the rods at a volumetric flow rate of 70 ml/min. Although the flow may not be fully developed, over the region that we captured, we did not detect any apparent evolution in the concentration distribution with axial distance *x*. In the coronal (x-z plane) and sagittal (x-y plane) images, viewing the concentration distribution over 9-10 rows of rods (excluding the edges of the sagittal images), rod-to-rod variation was not observed and the concentration distribution appeared to be roughly periodic.

As can be seen in Fig. 6 (a,c,e,f), looking along the spanwise direction, the middle of the channel, near the centerline, shows higher concentration of particles than the edges of the channel, in agreement with Fig. 4(i) and (j). Such a particle migration effect was observed for a $\phi_b = 40\%$ suspension by [67] in a smooth, low aspect ratio channel. In the current system, a nonuniform spanwise distribution of particles likely develops in the smooth channel upstream of the porous medium. However, we notice several features of the concentration distribution that are unique to this model porous medium system. First, in Figure 6 (a-d), it is clearly apparent that the particle concentration is enhanced in the free-flow region compared to the porous region. The concentration is particularly low at the bottom of the rods. Also, comparing Figure 6 (a vs. c; b vs. d), we notice that the concentration inside the porous structure is higher between the rods (c,d) compared to on top of the rods (a,b). The development from the initially uniform particle distribution to this nonuniform pattern indicates that particle migration from the channel edges to the centerline occurs both over and among the rods Therefore, the existence of the rods alters the particle migration in the channel, meaning that the particles move away from the rods and sort between the rods, making the maximum concentration generally in the free-flow region rather than inside the rods, where the center height of the channel lies. This effect can be observed specifically in Fig. 6(e,f) where the particles form bands in the free-flow region with the maximum particle concentration in the mid-center of the channel. Our experimental data suggest that, for concentrated suspensions, the particles experience the rods as narrower smooth walls where the majority of particles flow through the gaps between columns of rods. Looking along the spanwise direction, the bands observed in the free-flow region (Fig. 6e) roughly correspond to the five open gaps between the six columns of rods (Fig. 6a,f). The streamwise gaps between rods contain particles at average or higher than average concentration, but the velocity at those locations is quite low.

We also observe that regions of high particle volume fraction and low shear rate magnitude (e.g., $\dot{\gamma}_{xz}$ in Fig. 6 (a) and (c)) particularly in the free-flow region, coincide and the particle concentration profiles are maximum at the channel center-line. For instance, as can be observed in Fig. 4(i, j) inside the free-flow region where the velocity is maximum near the side walls, the concentration is lower, while the concentration is higher in the channel mid-center. In addition, for high bulk concentration suspensions, particle bands of high concentration occur near gaps between columns of rods, where the local channel depth effectively increases. As mentioned before, these results are then consistent with the concept of the particle stress balance for the pressure-driven flow of suspensions proposed in the suspension balance model by [17, 22]. In other words, the need for the suspension normal stress to be constant in the directions perpendicular to the mean motion, where the suspension is nearly at steady state and nearly at a fully developed state, results in migration of particles throughout the channel and inside the porous region in inhomogeneous ways. We further observe that particle concentration nonuniformity increases with the bulk suspension concentration, as can be seen in Fig. 7 and Fig. 8 (i.e., in y-z and x-z planes) that compare the concentration maps for 10%, 20%, 30%, and 40% suspensions. For example, by comparing Figs. 7 (a,b) and Figs. 7 (g,h), it can be clearly seen that for $\phi_b = 10\%$ the concentration distributions are nearly uniform, while the concentration maps reveal significant heterogeneity in the concentration distributions for $\phi_b = 40\%$. These results indicate that particle interaction effects, rather than an external force such as gravity, are driving the particles into increasingly nonuniform concentration distributions as the bulk concentration increases. It appears that the suspension particles redistribute along the height (*y*) direction so that the concentration, and therefore the suspension viscosity, is lower in the porous region, where the resistance to flow is greater due to the excluded volume of the rods and regions of high shear at the surfaces of rods. In addition, it appears that particles redistribute along the spanwise (z) direction away from the end walls and toward local regions of lower shear in the gaps between rods.

Fig. 8 further clarifies the porous media model impact and the structure of the rods on particle banding as particle concentration increases from 10% to 40%. No banding is observed at 10% or 20% bulk concentration, but bands become increasingly visible at 30% and then 40% bulk concentration. In general, our data in this study and from PIV studies of the same system [43] suggest that, for lower bulk suspension concentrations, the particles flow around each rod in the porous medium as an individual cylinder, while for higher concentrations, the particles tend to flow around each whole column of rods, as if there were a comb-shaped manifold channel, leading to distinct particle migration in the channel over and through the porous structure at low and high bulk concentrations.

Fig. 9 (a) represents the streamwise average value of the particle volume fraction obtained from Figure 8 against the spanwise (*z*) position. While the error bars are not shown for clarity, the uncertainty (standard deviation) was consistent for each bulk volume fraction at an average value of 0.06. The effect of the porous media rods structure on the particle banding in the free channel region for higher suspension concentrations can be further confirmed in Fig. 9 (a). No clear structure is formed for $\phi_b$= 10% and 20%, but local maxima and minima are observed for $\phi_b$= 30% and 40%.

Fig. 9 (b) and (c) show the streamwise velocity profile on top of rods and between the rods, obtained from Fig. 4, plotted against the spanwise direction. The streamwise velocity from Fig. 4 was averaged over the 4 adjacent pixels capturing the free flow region for Fig. 9 (b-c), matching the region plotted in Fig. 9 (a). In Fig. 9 (d), the velocity values from on top of rods and between the rods were averaged, and the average velocity profile is shown. These results show that another considerable difference between the Newtonian and suspension velocity field is the velocity variation with the span. The Newtonian suspending fluid and less concentrated suspensions show five approximately equal velocity peaks, corresponding to the gaps in the rods, while the 30% and 40% bulk concentration suspensions have lower peaks in the center and higher peaks near the two ends of the channel width. Error bars are not shown due to the overlapping profiles, but 0.3-0.4 cm/s are typical standard deviation values of the average profile.

Concentration profiles in Fig. 9 (e-g) were obtained from the axial MRI images in Figure 7, where the particle volume fraction was averaged over the 4 adjacent pixels capturing the free flow region and plotted against the spanwise direction ($z$ direction), for (e) concentration on top of the rods, (f) concentration between the rods, and (g) averaged values of both on top and between the rods. Here we see roughly uniform particle volume fraction at 10% and 20% bulk volume fraction and then multiple peaks forming at higher concentration, where the highest peaks are near the center and the peaks near the ends are lower. Error bars are not shown for readability, but the fluctuations between neighboring data points give a representative view of the uncertainty.

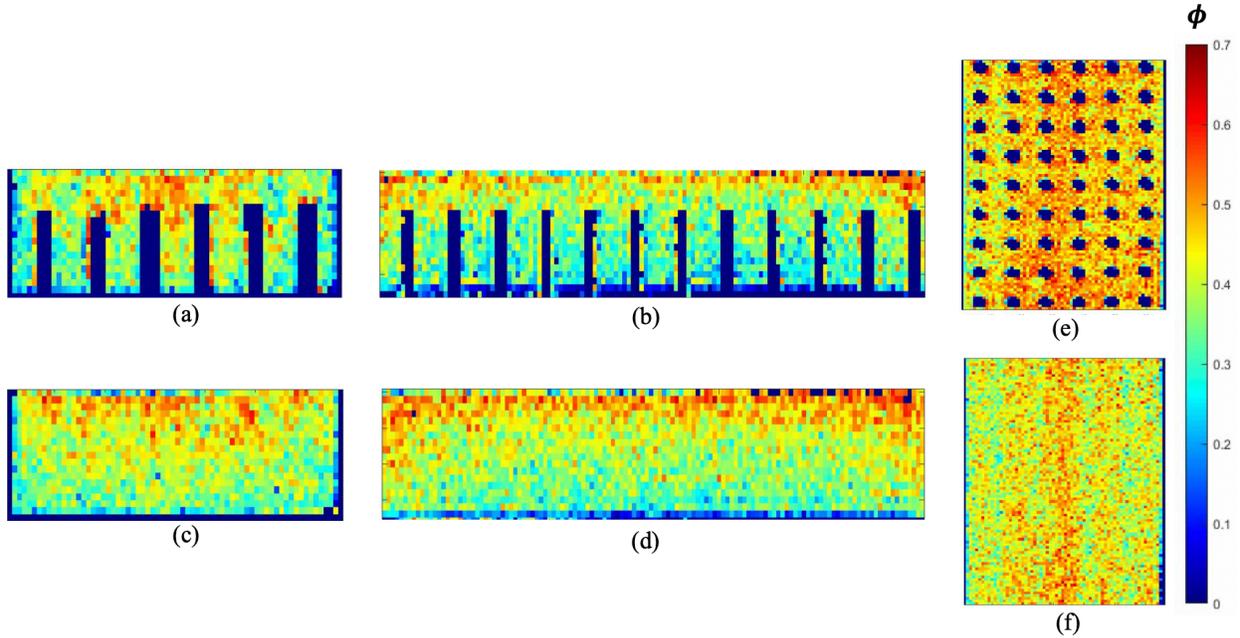

**FIGURE 6.** Particle concentration maps for $\phi_b = 40\%$ and Q=70 ml/min, (a) flow on top of the rods and (c) flow in between the rods for front view at (y-z plane at x=23.2 ± 0.11 cm), height of the flow cross-section (region shown) is 7 mm and the width is 2.5 cm, (b) flow on top of the rods and (d) flow between the rods for side view at (x-y plane at z=0 ± 0.11 cm), height of the flow cross-section (region shown) is 7 mm and the length is 5 cm, and (e) flow at the suspension-rods interface (x-z plane at y = $h_p$ - 0.5 mm = 4.5 mm) and (f) flow inside the free-flow region for top view at (x-z plane at y=$h_p$+0.5 mm = 5.5 mm), length of the flow cross-section (region shown) is 3.5 cm and the width is 2.5 cm. MRI scan parameters: spin-echo 2D pulse sequence, TR=2 s, TE=13.7 ms, field of view=5 cm × 5 cm, slice thickness=1 mm,128 × 128 pixels, 1 excitation, scan time=4 min. 16 sec.

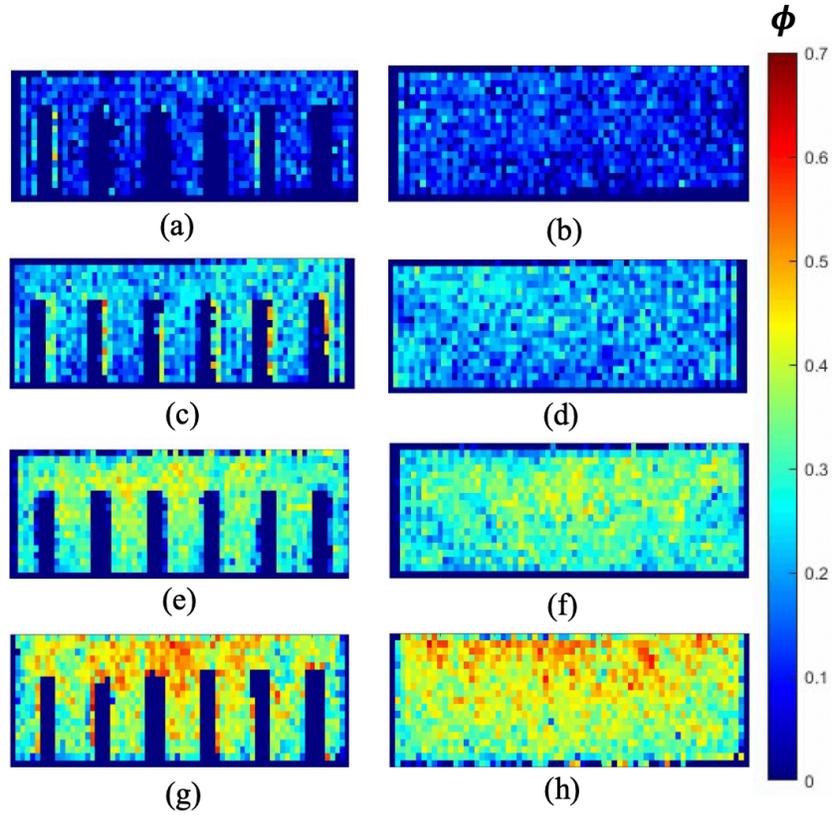

**FIGURE 7.** Particle concentration maps for (a,b) $\phi_b$ = 10%, (c,d) $\phi_b$ = 20%, (e,f) $\phi_b$ = 30%, and (g,h) $\phi_b$ = 40% for Q = 70 ml/min, (a,c,e,g) flow on top of the rods and (b,d,f,h) flow in between the rods for axial cross section view (y-z plane at x=23.2 + 0.11 cm). The height of the flow cross-section (region shown) is 7 mm and the width is 2.5 cm. MRI scan parameters: spin-echo 2D pulse sequence, TR=2 s, TE=13.7 ms, field of view=5 cm × 5 cm, slice thickness=1 mm, 128 × 128 pixels, 1 excitation, scan time=4 min. 16 sec.

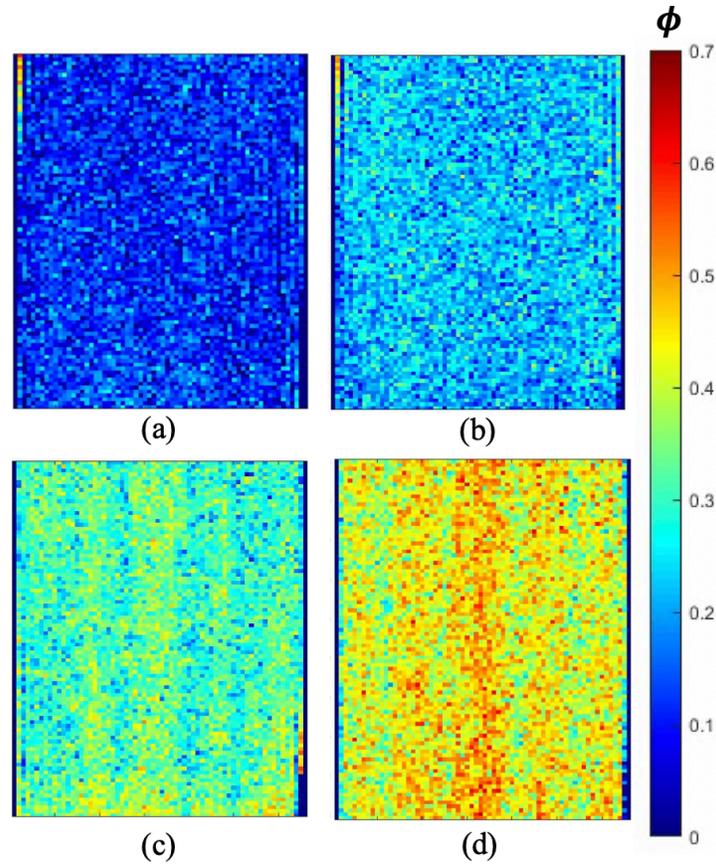

**FIGURE 8**. Particle concentration maps for (a) $\phi_b$ = 10%, (b) $\phi_b$ = 20%, (c) $\phi_b$ = 30%, and (d) $\phi_b$ = 40% for Q = 70 ml/min, flow inside the free-flow region for top view at (x-z plane at y=$h_p$+0.5 mm = 5.5 mm). The length of the flow cross-section (region shown) is 3.5 cm and the width is 2.5 cm. MRI scan parameters: spin-echo 2D pulse sequence, TR=2 s, TE=13.7 ms, field of view=5 cm × 5 cm, slice thickness=1 mm, 128 × 128 pixels, 1 excitation, scan time=4 min. 16 sec.

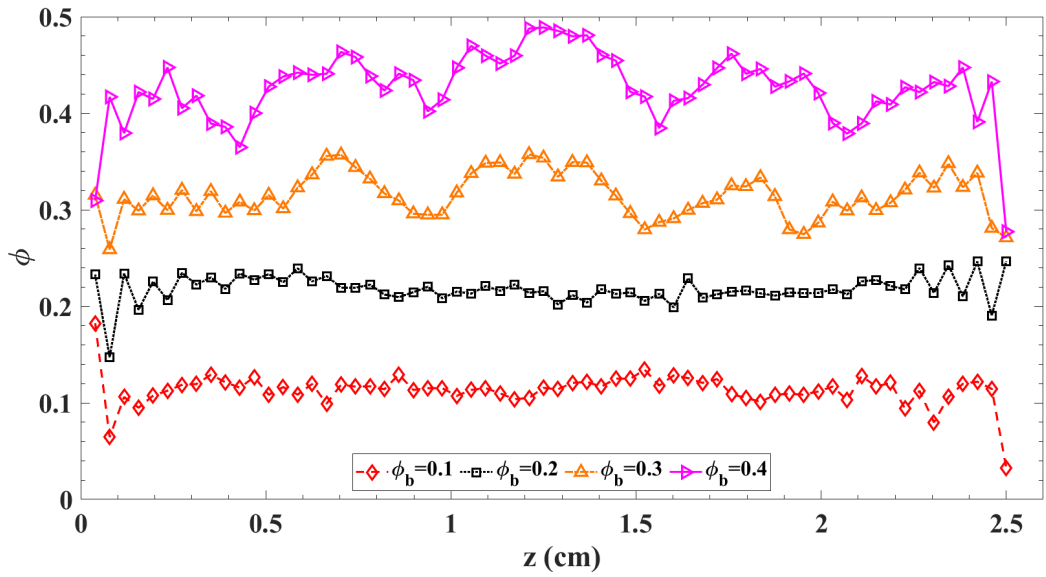

(a)

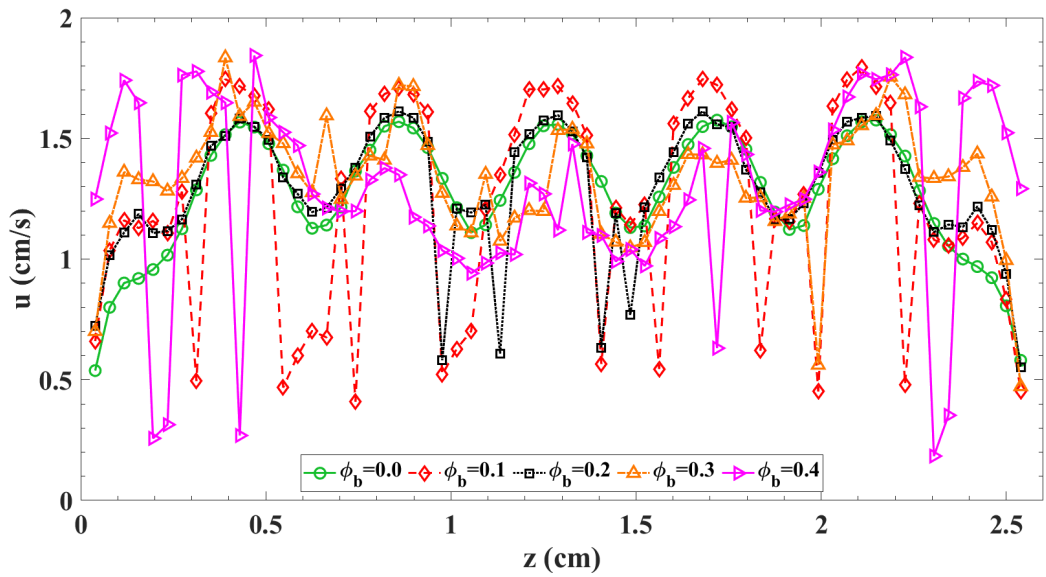

(b)

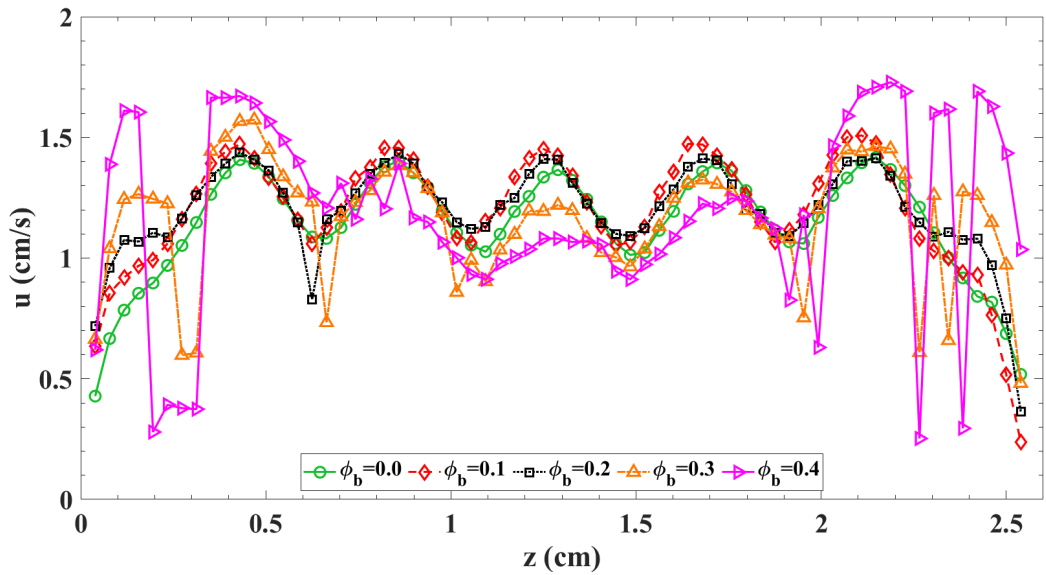

(c)

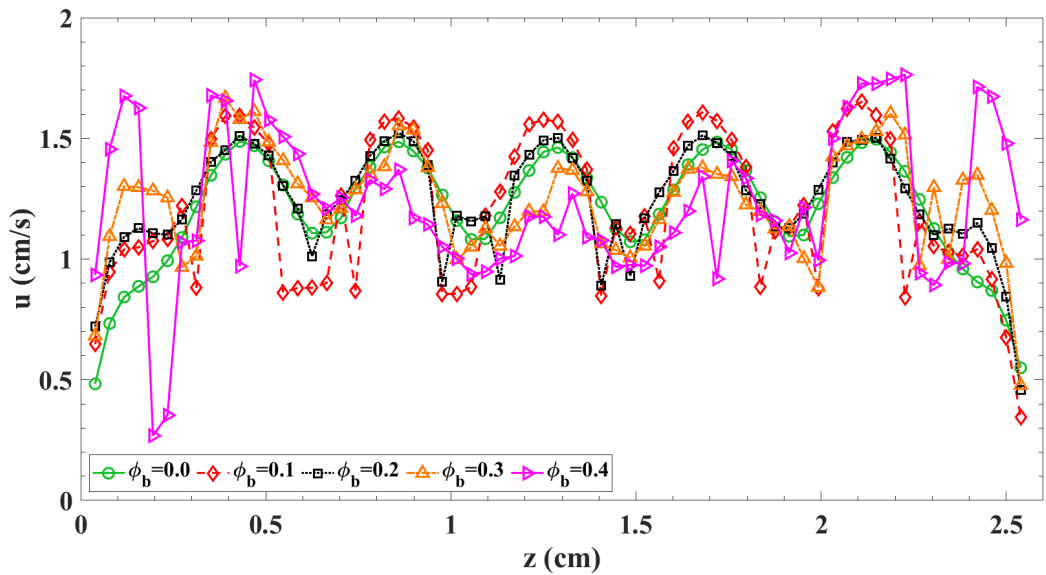

(d)

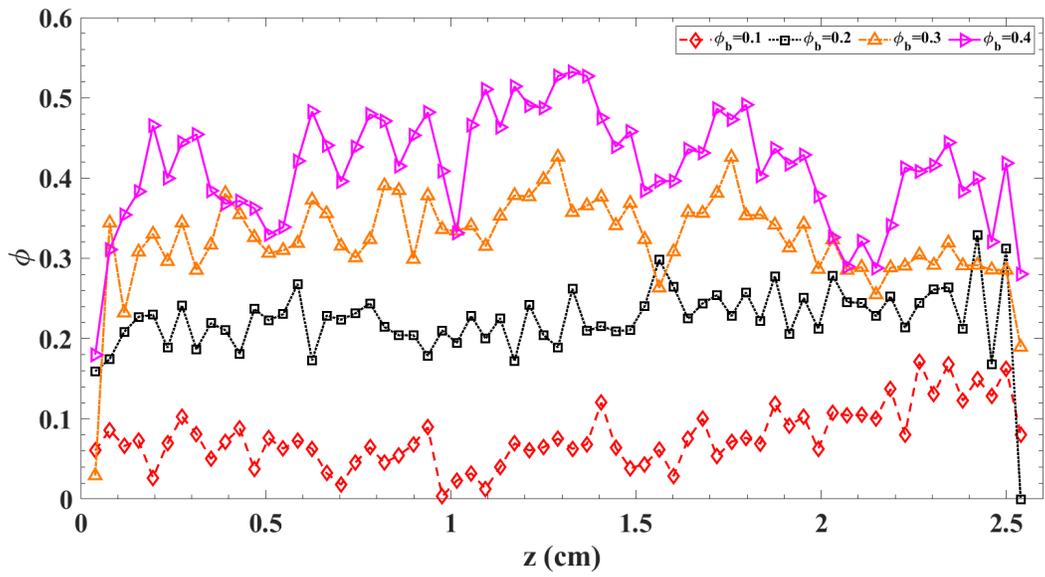

(e)

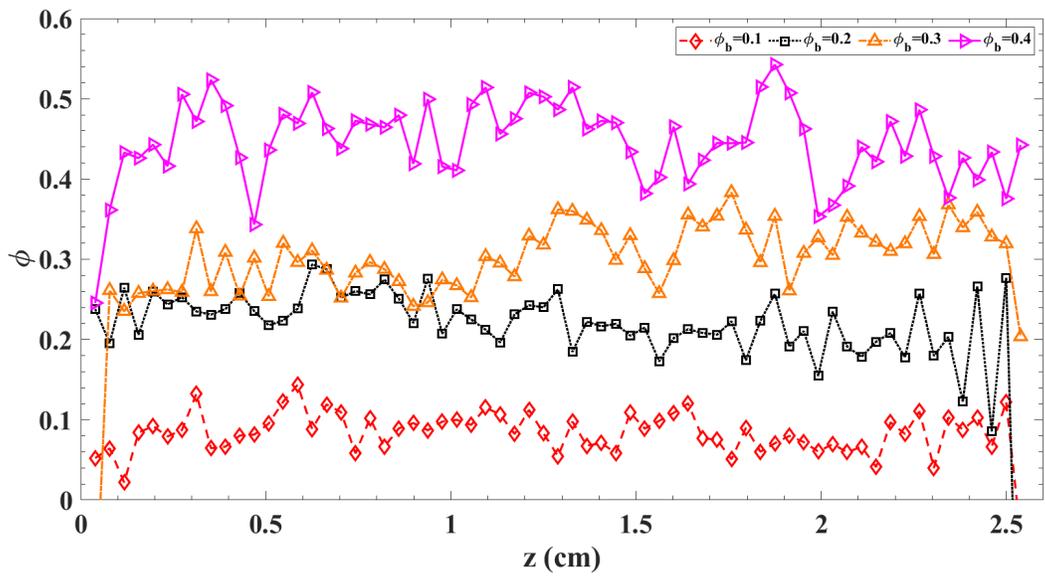

(f)

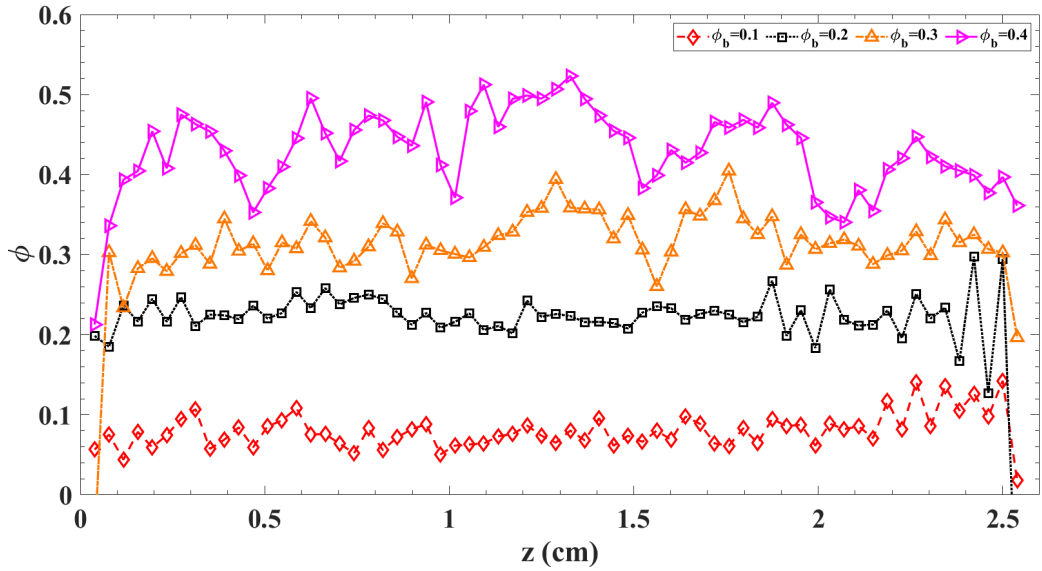

(g)

**FIGURE 9.** Particle concentration and velocity profiles along the channel width in the spanwise direction ($z$ direction), for bulk concentrations ranging from 10% to 40%. The profile in (a) was obtained from the MRI top view concentration maps shown in Figure 8 (x-z plane at $y=h_p+0.5$ mm = 5.5 mm), where the values in the map for each bulk concentration were averaged along the axial length of the cross-section shown (3.5 cm). Velocity profiles in (b-d) were obtained from the axial MRI images in Figure 4, where the streamwise velocity was averaged over the 4 adjacent pixels capturing the free flow region and plotted against the spanwise direction ($z$ direction), for (b) flow on top of the rods, (c) flow between the rods, and (d) averaged values of both on top and between the rods. Concentration profiles in (e-g) were obtained from the axial MRI images in Figure 7, where the particle volume fraction was averaged over the 4 adjacent pixels capturing the free flow region and plotted against the spanwise direction ($z$ direction), for (e) flow on top of the rods, (f) flow between the rods, and (g) averaged values of both on top and between the rods. The volumetric flow rate was $Q = 70$ ml/min for all cases.

Fig. 10 (a, b, c) shows the particle concentration profile along the channel height ($y$ direction) on the channel centerline ($z=0$) for various bulk flow concentrations ranging from 10% to 40% on top of the rods, between the rods, and averaged values of both on top and between the rods, respectively. Each plot shows the concentration values averaged over the 3-4 most central pixel columns from the axial concentration maps of Figure 7. For simplicity, we deleted the error bars of the standard deviation in the plot. For each bulk concentration, the average standard deviation of the measured particle volume fraction was close to 0.06. For $\phi_b = 10\%$, the particle concentration is relatively uniform across the channel height. As the concentration increases to 20%, 30% and 40%, the higher concentration occurs in the free-flow region, whereas the porous region has low particle concentration. This asymmetry with maximum concentration in the free-flow region can also be observed in the axial and sagittal concentration maps shown in Figs. 6(a-d).

To further characterize the suspension flow inside the porous structure, we also plot the average particle flux fraction and the average flow fraction through the porous model for various suspension concentrations and Q=70 ml/min (see Fig. 11). The flow fraction was calculated from the axial velocity maps in Figure 4. The portion of the volumetric flow rate passing through the porous region ($y \leq 0$) was summed over the entire axial cross section and divided by the total volumetric flow rate, for the image on top of the rods and the image between the rods and then the two values were averaged to yield the average flow fraction. The particle flux fraction was calculated from the axial concentration maps in Figure 7. The particle flux in each pixel of the image was calculated by multiplying the local velocity by the local concentration. The portion of the particle flux passing through the porous region ($y \leq 0$) was summed over the entire axial cross section and divided by the total particle flux, for the image on top of the rods and the image between the rods and then the two values were averaged to yield the average particle flux fraction. In general, the flow fraction and particle flux fraction decay slightly, on the order of 5%, as the concentration increases. The data suggest that the flow fraction and particle flux fraction are relatively constant over the examined range of bulk concentrations. Perhaps the observed flow fraction and particle flux fraction correspond to a minimum value of the pressure drop or viscous dissipation. Differences between the particle flux fraction and flow fraction might be due to particle banding at high concentration. It should be noted that porous structure and thickness can impact these results.

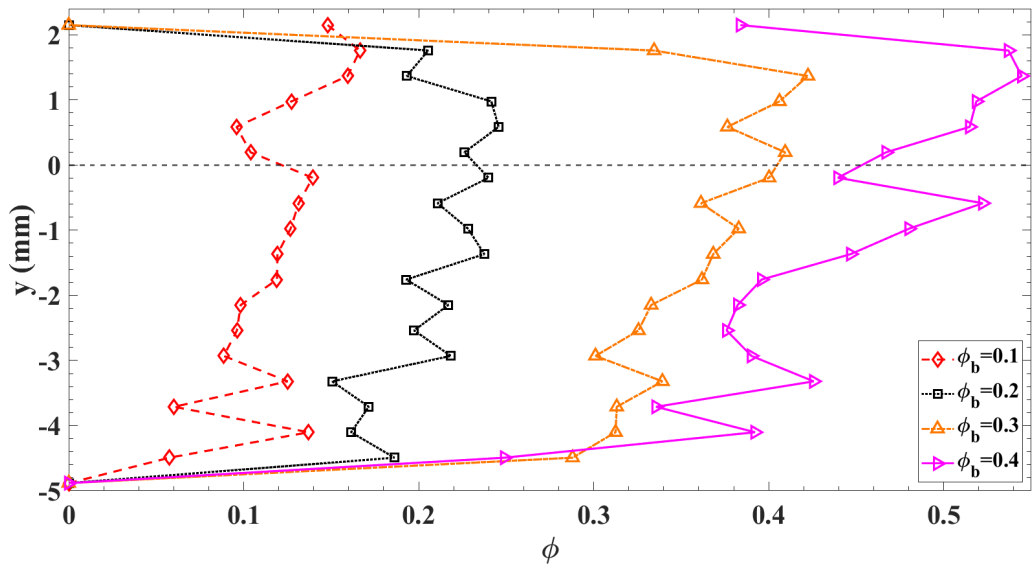

(a)

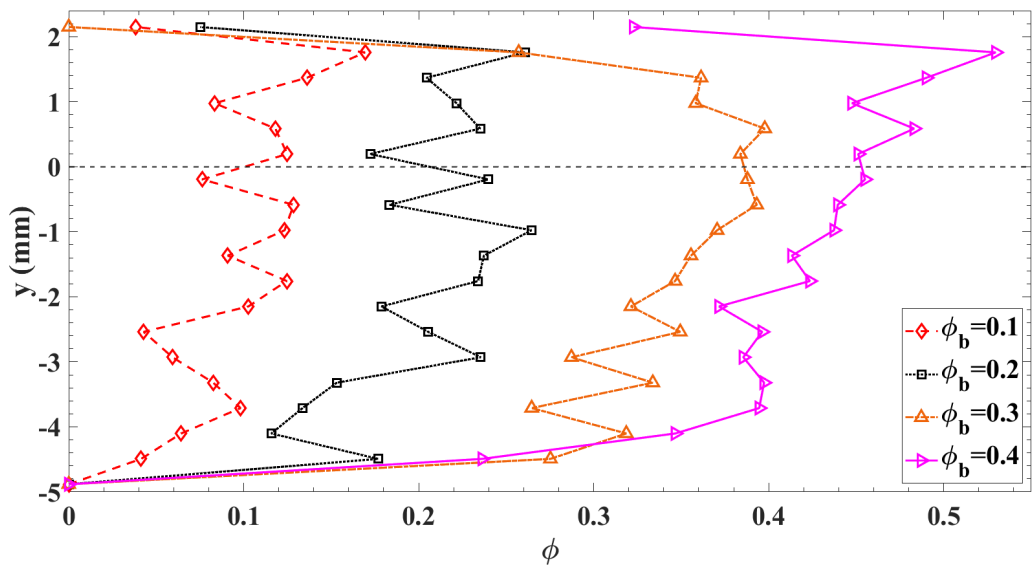

(b)

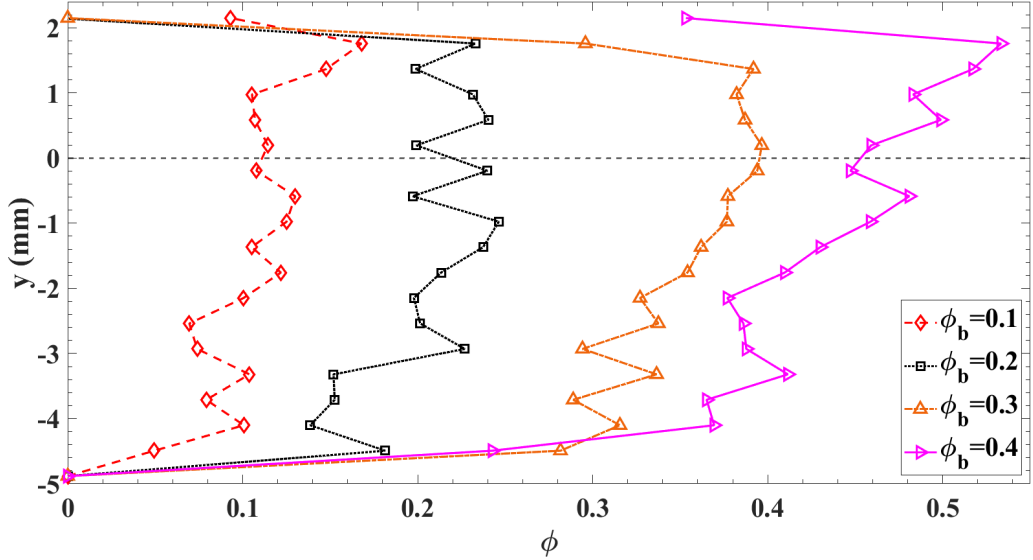

(c)

**FIGURE 10.** Particle concentration profiles along the channel height ($y$ direction) obtained from the MRI axial concentration maps shown in Figure 7, for (a) flow on top of the rods, (b) flow between the rods, and (c) averaged values of both on top and between the rods, averaging over the 3-4 adjacent pixels on the centerline of the channel for bulk concentrations ranging from 10% to 40%. The data on the rods and between the rods of the porous structure were averaged. The volumetric flow rate was $Q = 70$ ml/min for all cases.

## 4. Conclusion

We investigated the suspension flow behavior of non-Brownian, non-colloidal suspensions in a channel using MRI measurements where the bottom surface of the channel is covered with a model porous structure consisting of an array of cylindrical rods. Our specific focus was characterizing particle concentration maps, velocity maps and corresponding profiles due to the existence of the structured surface. A series of experiments was performed, for the first time, for particle bulk volume fractions ranging from 10% to 40% while the porous structure had known material properties (i.e. porosity and permeability) and thickness.

While the velocity profiles for low concentrations of suspensions (i.e., 10% and 20%) appear to be similar to those of the pure suspending fluid, as the suspension concentration increases, due to the particle-particle, particle-rod, and particle-wall interactions, the velocity profile is modified. At higher bulk particle concentrations of 30% and 40%, we observe nonuniform particle distributions due to the action of shear-induced particle migration. Particle concentration is enhanced in the free-flow region above the porous medium and reduced at the bottom of the rods in the porous medium. Also, by increasing the concentration, the velocity through the porous region decreases. The porous structure and the rods' arrangement also play a critical role in the flow. In this case, the structure of the rods also modifies particle migration inside the porous region such that by adding more particles, they tend to accumulate slightly in the y-z axial planes between rods where the shear rate is lower compared to the axial planes on top of rods. It should be noted that in impermeable geometries while at low Reynolds number, particles in dilute suspension flows follow streamlines, whereas at higher concentrations, hydrodynamic interactions between particles lead to self-diffusion in uniform flows and shear-induced migration in non-uniform shear flows [80-82]. The heterogeneous concentration fields and modified velocity profiles as a result of shear-induced migration have also been reported in (e.g., [12, 18, 83, 84]). One of the critical outcomes of this study is the particle banding effect due to the existence of gaps between columns of rods in the porous media model. It appears that the surface structure leads to the migration of particles to specific locations, leading to banding, in the vicinity of the free-flow region, especially as the volume fraction increases. This impact depends on various parameters including the suspension properties, particle size and shape, and also the thickness and the properties of the porous structure. Although the present experiments are all conducted for $\frac{H-h_p}{a} \approx$ 24 (i.e., continuum effect conserved), our recent numerical simulations [85], showed that banding occurs in a Couette flow where the surfaces are covered with porous media even though the particles are not moving inside the porous media.

Our study quantifies the impact of a structured surface on the flow of non-colloidal and non-Brownian suspensions in the low Reynolds number regime. Indeed, it would be interesting to vary these porous medium parameters and to further investigate their impact on particle migration, and the resulting heterogeneous concentration fields and modified velocity profiles. We expect that the present experimental data could be used; 1) to further improve the continuum models such as DFM and SBM to capture the particle migration over and through structured surfaces, 2) to further investigate the impact of both particles and structures in the flow using discrete-particle simulations, and 3) to improve equations related to the porous media such as Volume Average Navier-Stokes (VANS) equations in order to consider the impact of particles in the model analysis.

It would be interesting to investigate geometric effects in the future including two-dimensional grooved geometries.

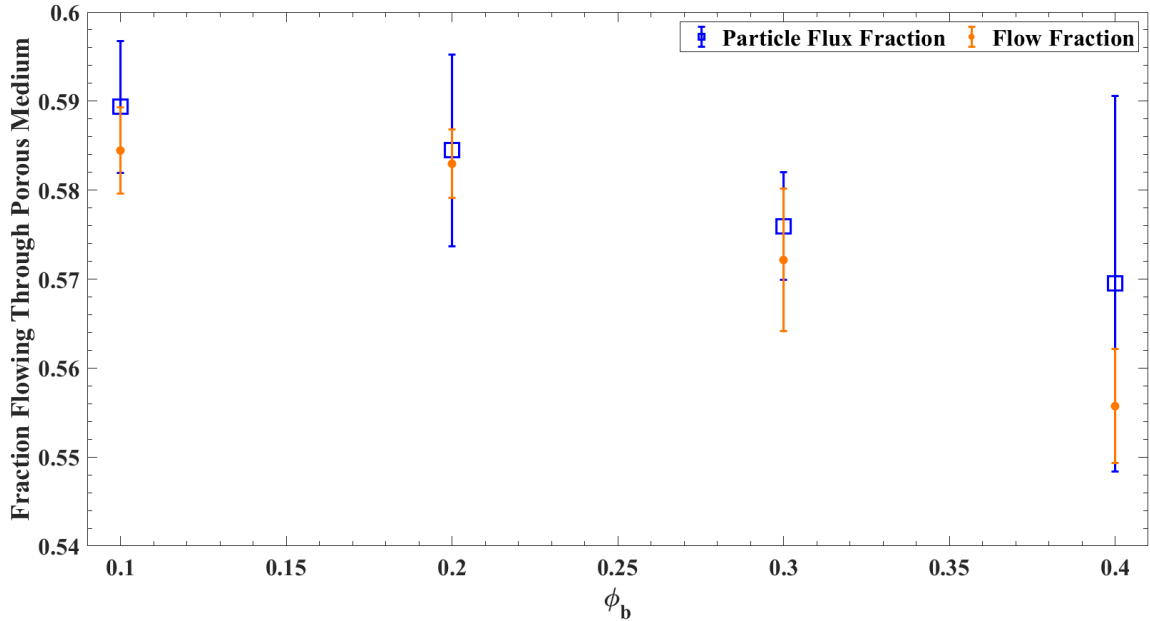

**FIGURE 11.** Average flow and particle flux fractions inside the porous model vs. the bulk particle volume fraction, ranging from 10% to 40% for Q=70 ml/min.

**Acknowledgements**

This work has been supported partially by National Science Foundation award #1854376 and partially by Army Research Office award #W911NF-18-1-0356. The authors also gratefully acknowledge the Rutgers University Molecular Imaging Center for aiding with MRI scanning.